\documentclass[sigconf]{acmart}

\usepackage{enumitem}
\usepackage{tikz}

\usepackage{filecontents}

\usepackage{algorithmic}
\usepackage{graphicx}
\usepackage{textcomp}
\usepackage{url}

\usepackage[normalem]{ulem}

\usepackage{subfig}
\usepackage[normalem]{ulem}
\usepackage{comment}
\usepackage{algorithmic}
\usepackage{graphicx}
\usepackage{textcomp}
\def\BibTeX{{\rm B\kern-.05em{\sc i\kern-.025em b}\kern-.08em
    T\kern-.1667em\lower.7ex\hbox{E}\kern-.125emX}}

\usepackage{tikz}
\definecolor{aqua}{rgb}{0.0, 1.0, 1.0}
\definecolor{teal}{rgb}{0.0, 0.5, 0.5}
\definecolor{maroon}{rgb}{0.5, 0.0, 0.0}
\definecolor{lime(web)(x11green)}{rgb}{0.0, 1.0, 0.0}

\usepackage{pifont}
\usepackage{hyperref}
\usepackage{lscape}
\usepackage{numprint}
\usepackage{array}
\usepackage{multicol}
\usepackage{longtable}
\usepackage{mathtools}
\usepackage{caption}
\usepackage{makecell}
\usepackage{tikz}

\usepackage{pgfplotstable}

\copyrightyear{2023}

\acmConference[CAMS '23]{CAMS '23: Workshop on Computer Architecture Modeling and Simulation (with MICRO 2023)}{October 28, 2023}{Toronto, Canada}
\setcopyright{iw3c2w3}

\begin{document}

\newcommand{\note}[1]{\{\textcolor{blue}{#1}\}}

\title{Accelerating Computer Architecture Simulation through Machine Learning}

%

\author{Wajid Ali}
\email{malikwajid441@gmail.com}
\affiliation{%
  \institution{Dept. of ECE, UET Lahore}
  \state{}
  \country{}
}

\author{Ayaz Akram}
\email{yazakram@ucdavis.edu}
\affiliation{%
  \institution{Dept. of CS, UC Davis}
  \state{}
  \country{}
}

\date{}

\renewcommand{\shortauthors}{Ali and Akram}

\begin{abstract}
    This paper presents our approach to accelerate computer architecture simulation by leveraging machine learning techniques. Traditional computer architecture simulations are time-consuming, making it challenging to explore different design choices efficiently. Our proposed model utilizes a combination of application features and micro-architectural features to predict the performance of an application. These features are derived from simulations of a small portion of the application. We demonstrate the effectiveness of our approach by building and evaluating a machine learning model that offers significant speedup in architectural exploration. This model demonstrates the ability to predict IPC values for the testing data with a root mean square error of less than 0.1.
\end{abstract}

\maketitle

\section{Introduction}
\label{sec:intro}

Computer architecture simulation stands as a cornerstone in the quest to design and optimize efficient computing systems~\cite{Akram2016}. Traditionally, simulators like gem5~\cite{lowe2020gem5} have been instrumental in providing detailed insights into the intricate interactions between software applications and hardware microarchitectures. However, these simulations often demand significant computational resources and time, constraining the pace of architectural exploration.

In this context, our research seeks to augment traditional simulators like gem5 with the power of machine learning. Our vision revolves around the development of a machine learning model that mimics gem5's behavior, allowing users to harness the synergy between simulation and machine learning based prediction. This paradigm shift envisions a scenario where users simulate an application for a brief time interval and leverage the resulting simulation data to predict the application's performance over more extended time periods.

\subsection{Advancing Simulation with Machine Learning}

Machine learning techniques have already been used for performance prediction in various contexts~\cite{akram2020tribes, ardalani2015cross, li2009machine}.~Our endeavor finds inspiration in similar approaches that have sought to enhance the accuracy and efficiency of simulators. Renda et al.~\cite{renda2020difftune}, for instance, paved the way by exploring the integration of machine learning techniques into architectural simulation. 

\subsection{Two Pathways to Machine Learning Integration}

The first pathway centers on the collection of application statistics that are microarchitecture-independent and readily obtainable on real machines or functional simulators and emulators, known for their speed. These statistics encompass characteristics such as the types of instructions within the application code. By training machine learning models using these features, we can create models that are not influenced by microarchitectural intricacies. However, such models, while potentially swift and versatile, may lack the flexibility to account for the nuances of the underlying hardware.

The second pathway explores the incorporation of microarchitectural features into the machine learning model. This approach endows the model with the capability to adapt and learn the nuanced impact of hardware-specific attributes. By considering both application-specific statistics and microarchitectural features, we endeavor to construct a more comprehensive model capable of capturing the complex interactions between software and hardware.

\subsection{Our Research Objectives}

Our primary objective is to model the behavior of gem5, a well-established and versatile architectural simulator. Our vision is to empower more detailed simulations with the integration of a machine learning model, thus accelerating the simulation process. By achieving this goal, we aim to offer architects and researchers a swifter and more insightful means of investigating various design choices and their implications on application performance.

In the ensuing sections of this paper, we delve into the methodology, experiments, and results that constitute our exploration of these two machine learning integration pathways. Through empirical evidence, we hope to pave the way for a future computer architecture simulation research, one marked by a harmonious interplay between traditional simulation and machine learning-enhanced prediction.
\section{Methodology}
We conducted extensive simulations to collect data for our machine learning-based performance prediction model. This section details our simulation setup and data collection process.

We focused our simulations on the out-of-order CPU model of gem5~\cite{lowe2020gem5}, specifically the O3CPU model. This CPU model is widely used in computer architecture research and provides a suitable basis for our experiments. The configuration details of the O3CPU model based systems we rely on are presented in Table~\ref{tab:Config} for reference.

To ensure the relevance and diversity of our dataset, we selected high-performance workloads for our simulations. Specifically, we utilized the NAS Parallel Benchmark suite (NPB)~~\cite{bailey1991parallel}  and graph workloads from the GAPBS~\cite{beamer2015gap} suite. The NAS Parallel Benchmark suite consists of various computational kernels and pseudo applications designed for high-performance computing (HPC) systems. On the other hand, the GAPBS suite offers a collection of graph processing kernels.

To obtain a comprehensive dataset, we performed multiple simulation runs for each workload. For all configurations listed in Table~\ref{tab:Config}, we conducted simulations using three core configurations and three different DRAM devices. This resulted in a total of nine unique configurations.

We executed these simulations over varying time intervals to capture a wide range of dynamic behaviors exhibited by the applications. This extensive data collection effort allowed us to observe how different configurations and simulation intervals affected the applications' performance.


The outcomes of these simulations served as the training data for our machine learning models. By capturing the performance of the applications under different architectural conditions, we aimed to create predictive models that could estimate performance accurately and rapidly.

In summary, our methodology involved running simulations of high-performance workloads on the gem5 simulator, specifically utilizing the O3CPU model. We systematically varied configurations and simulation intervals to ensure a diverse dataset. This dataset, comprising performance metrics and corresponding feature vectors, formed the basis for training our machine learning models, as elaborated in subsequent sections.

\begin{table}[!h]
    \begin{center}
    \caption{\label{tab:Config}System Configuration Used for Experiments}
    \begin{tabular}{|p{2.45cm}|p{1cm}|p{1.3cm}|p{1cm}|}
      \hline
      Features & Baseline & Aggressive & Lean \\
      \hline
      Number of cores & 8 & 8 & 8 \\
      Core type & OoO & OoO & OoO \\
      Core width & 8 & 16 & 4 \\
      ROB entries/core & 192 & 384 & 96 \\
      \hline
      Private L1 Inst. & 32 KB & 64 KB & 16 KB \\
      Private L1 Data & 512 KB & 1024 KB & 256 KB \\
      Shared L2 & 8 MB & 16 MB & 4 MB \\
      \hline
      \multicolumn{4}{|c|}{Main Memory} \\
      \hline
      DRAM size & \multicolumn{3}{|c|}{128GiB} \\ \hline
      DRAM device & \multicolumn{3}{|c|}{DDR4\_2400, LPDDR5\_6400, DDR5\_6400}  \\
      \hline
    \end{tabular}
    \label{tab:conftable}
    \end{center}
    \end{table}

\section{Machine Learning Models and Input Features}

In this section, we detail the machine learning models employed in our study and the selection of input features, which are crucial for the performance prediction task.

\subsection{Input Features}

To facilitate the training of various machine learning models for performance prediction, we carefully selected a set of informative input features that capture critical aspects of the application and its execution on the simulated architecture. These features are presented in Table~\ref{tab:inpfeatures} along with their descriptions.

These features were selected based on their relevance to the performance characteristics of the applications and the microarchitecture. They provide valuable insights into instruction-level behavior, memory accesses, and cache utilization, all of which significantly influence application performance.

\subsection{Machine Learning Models}

To explore the efficacy of machine learning models in predicting application performance, we experimented with several widely-used models, including:

\textbf{Linear Regression:} Linear regression~\cite{james2013introduction, seber2003linear} is a fundamental regression technique that assumes a linear relationship between the input features and the target variable. It provides interpretable coefficients for each feature.

\textbf{Support Vector Machine (SVM):} SVM~\cite{boser1992training, james2013introduction} is a versatile supervised learning algorithm that aims to find a hyperplane that best separates the data points. In our context, we utilized SVM for regression tasks, seeking to establish a predictive model for application performance.

\textbf{Random Forest:} Random Forest~\cite{breiman2001random, james2013introduction} is an ensemble learning technique that combines multiple decision trees to make predictions. It is known for its robustness and ability to capture complex relationships in the data.

\subsection{Model Selection and Performance}

Among the experimented machine learning models, the Random Forest model emerged as the most effective in predicting application performance. In the subsequent section, we present and discuss the results obtained using the Random Forest model.

It is important to note that our goal was not only to determine the best-performing model but also to conduct a preliminary exploration of different machine learning models' suitability for performance prediction. This exploration provides valuable insights into the feasibility of utilizing machine learning techniques in computer architecture research.

In the following section, we delve into the results obtained from our experiments with the Random Forest model, showcasing its effectiveness in predicting application performance based on the selected input features.

\begin{figure}
    \centering
    \includegraphics[scale=0.42]{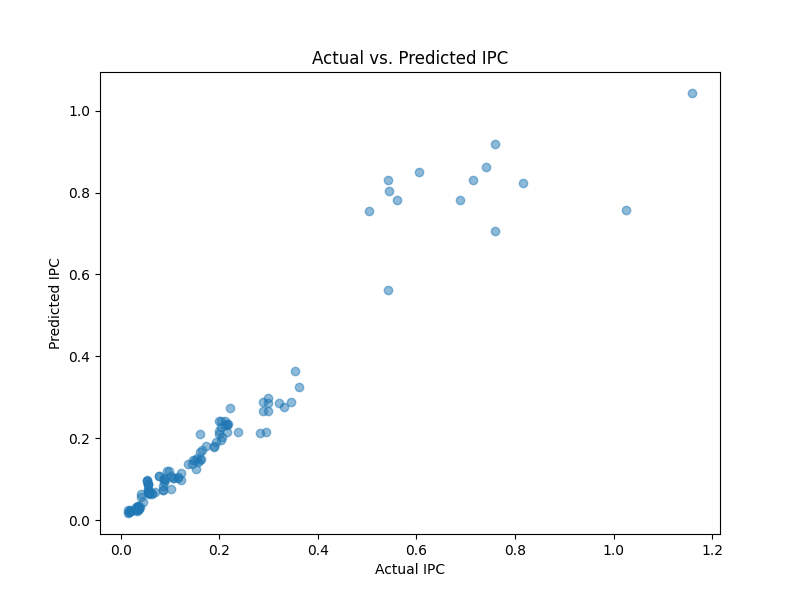}
    \caption{Predicted vs. Actual IPC}
    \label{fig:rforestpred}
  \end{figure}

  \begin{figure}
    \centering
    \includegraphics[scale=0.42]{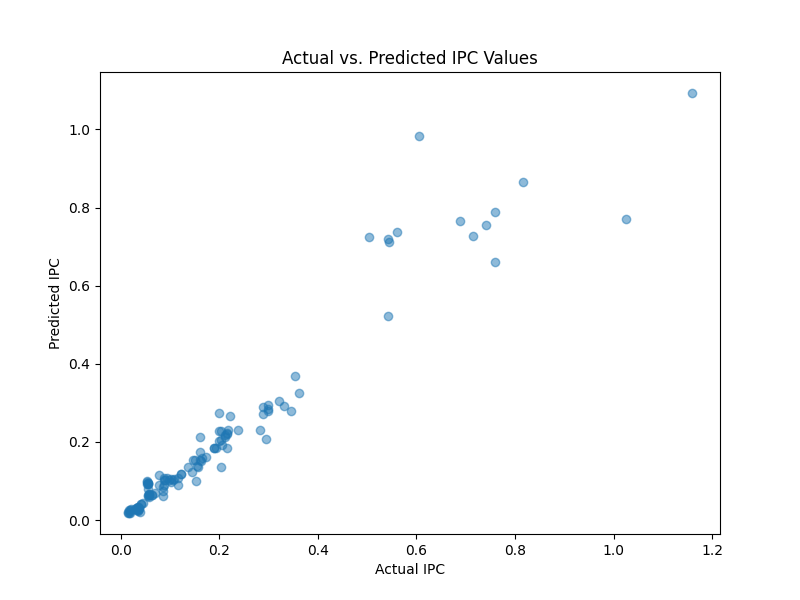}
    \caption{Predicted vs. Actual IPC when the training data is normalized to the instruction count.}
    \label{fig:rforestprednorm}
  \end{figure}

  \begin{figure*}
    \centering
    \includegraphics[scale=0.45]{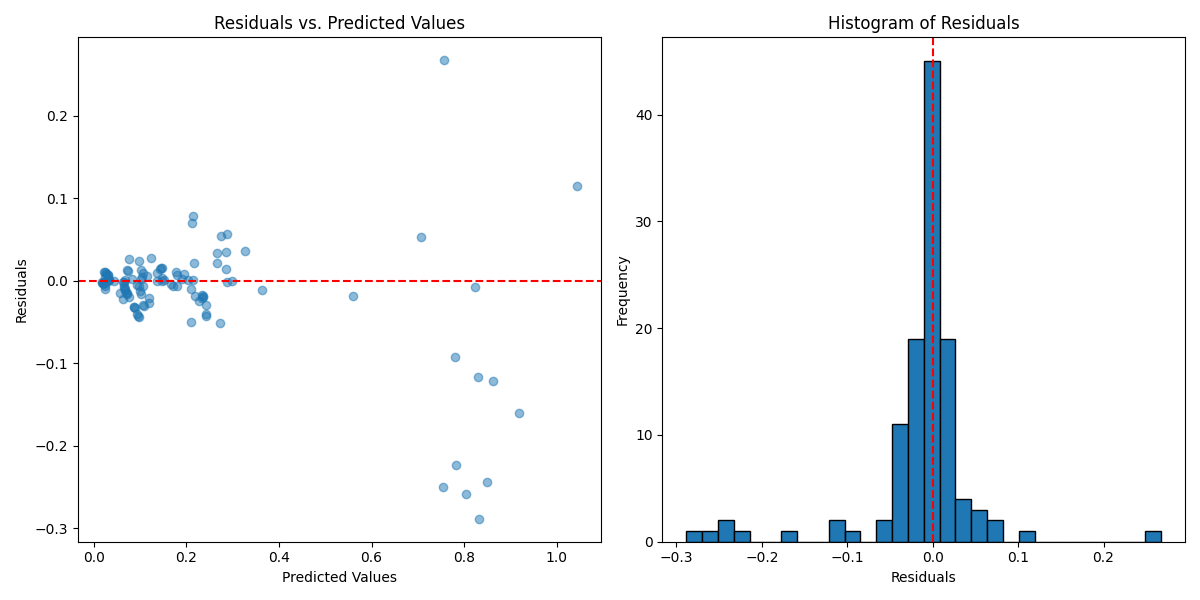}
    \caption{Residual analysis for IPC Prediction.}
    \label{fig:rforestresid}
  \end{figure*}

\begin{table}
    \centering
        \caption{Features Used in the Model}
        \begin{tabular}{|p{2cm}|p{6cm}|}
        \Xhline{2\arrayrulewidth}
        \textbf{Feature} & \textbf{Description} \\ \Xhline{2\arrayrulewidth}
        \textbf{numLoadInsts} & number of load instructions in the application \\ \hline
        \textbf{numStoreInsts} & number of store instructions in the application \\ \hline
        \textbf{numInsts} & total instructions in the application \\ \hline
        \textbf{numBranches} & number of branch instructions in the application \\ \hline
        \textbf{numOps} & number of micro-operations in the application \\ \hline
        \textbf{L1IcacheSize} & L1 instruction cache size \\ \hline
        \textbf{L1Dcache} & L1 data cache size \\ \hline
        \textbf{L2cache} & L2 cache size \\ \hline
        \textbf{pipelineWidth} & width of the processor's execution pipeline \\ \Xhline{2\arrayrulewidth}
    \end{tabular}
    \label{tab:inpfeatures}
\end{table}

%
%
%

\section{Results and Discussion}
\label{sec:results}




\begin{figure*}
  \centering
  \includegraphics[scale=0.45]{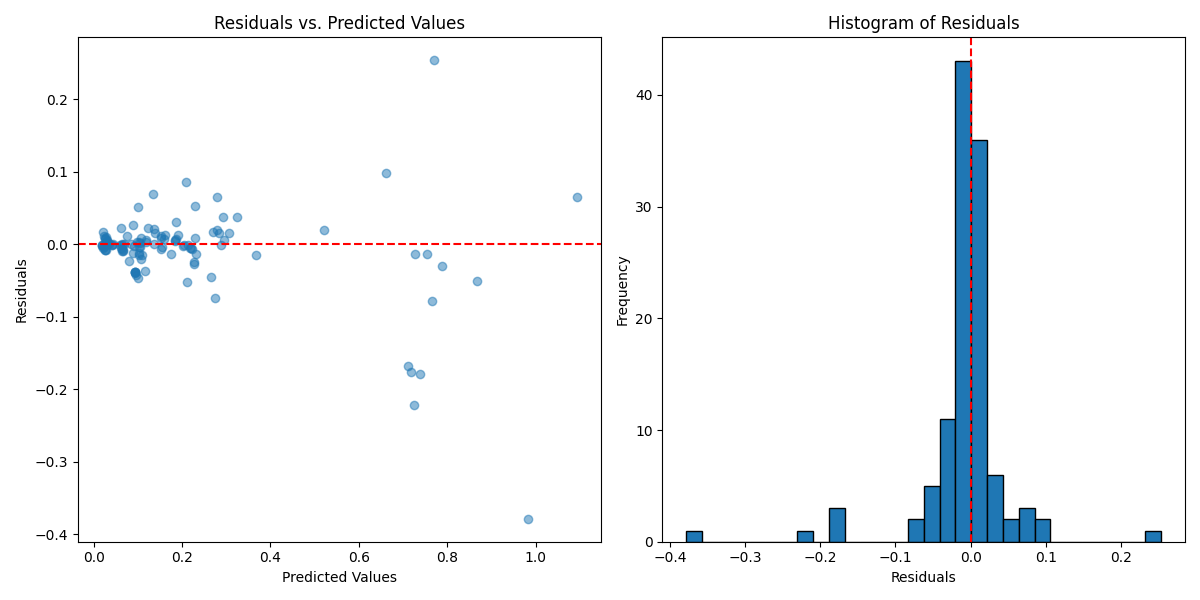}
  \caption{Residual analysis for IPC prediction when the training data is normalized to the instruction count.}
  \label{fig:rforestresidnorm}
\end{figure*}

\begin{figure}
  \centering
  \includegraphics[scale=0.55]{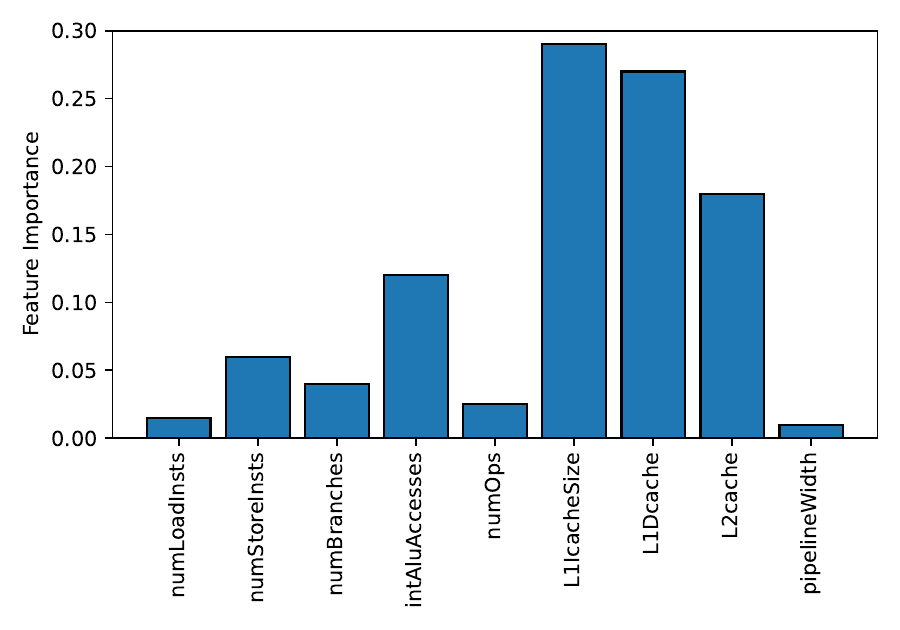}
  \caption{Feature importance for IPC prediction}
  \label{fig:featureimp}
\end{figure}

In this section, we present the outcomes of our machine learning-based performance prediction model. Our primary performance metric of interest was "instructions per cycle" (IPC). We evaluated the model's predictions on the testing data and assessed its accuracy.

\subsection{Predicted vs. Actual IPC}

Figure~\ref{fig:rforestpred} illustrates the relationship between the predicted IPC values generated by our Random Forest model and the actual IPC values observed in the test data. The close alignment between the predicted and actual IPC values is evident, demonstrating the model's effectiveness in approximating application performance.

\subsection{Root Mean Square Error (RMSE)}

To quantitatively evaluate the model's predictive accuracy, we computed the Root Mean Square Error (RMSE). The RMSE measures the average magnitude of prediction errors, with lower values indicating better predictive performance. Our model achieved a RMSE value of less than 0.1, signifying its ability to closely match the actual IPC values.

\subsection{Residual Analysis}

Figure~\ref{fig:rforestresid} displays the residual values obtained during IPC prediction. Residual values represent the differences between predicted and actual IPC values. A residual of 0 indicates perfect prediction, while small residual values indicate accurate predictions. As observed in Figure~\ref{fig:rforestresid}, the residual values are consistently small, reinforcing the model's reliability in estimating IPC.

\subsection{Normalized Performance Prediction}

To extend the utility of our model for longer simulation times while relying on a small simulation interval, we normalized our training data to the number of instructions in the program. This normalization allows for the prediction of IPC values over extended execution times. Figures~\ref{fig:rforestprednorm} and \ref{fig:rforestresidnorm} present the results of this normalized model.

The normalized model demonstrates consistent accuracy in predicting IPC values over longer simulation intervals, making it a valuable tool for architectural exploration that leverages small simulation intervals.

In conclusion, our machine learning-based performance prediction model, particularly the Random Forest variant, exhibited a remarkable ability to predict application performance in terms of IPC. The close match between predicted and actual values, low RMSE, and small residual values attest to the model's effectiveness. Additionally, the normalization of training data has extended the model's applicability for predicting performance over extended simulation intervals, enhancing its utility for computer architecture research and design exploration.

\subsection{Importance of Input Features}


Understanding the significance of input features in our machine learning model is crucial for gaining insights into the relationships between application behavior and microarchitectural characteristics. Figure ~\ref{fig:featureimp} offers a visual representation of the importance our model assigns to various input features during the IPC prediction task.

As depicted in Figure~\ref{fig:featureimp}, our machine learning model has learned to assign the highest importance to cache sizes when predicting IPC. This observation underscores the pivotal role that cache configurations play in influencing application performance. By emphasizing cache sizes, our model recognizes their impact on memory access patterns, data availability, and, subsequently, overall execution efficiency.

Moreover, this insight highlights the potential of machine learning in providing a deeper understanding of simulation data. By analyzing feature importance, we can discern the sensitivity of different configurations and applications to specific microarchitectural attributes, such as cache size. This newfound knowledge can guide architectural decisions and optimizations, shedding light on which aspects of the architecture have the most significant impact on performance.

In essence, our machine learning approach not only yields accurate performance predictions but also opens the door to enhanced comprehension of the complex relationships between application workloads and microarchitectural features. By recognizing the pivotal role of cache sizes, we underscore the value of machine learning as a tool for gaining valuable insights into computer architecture and design.

\section{Conclusion}
\label{sec:conclusion}
%

In this research, we embarked on a preliminary exploration to accelerate computer architecture simulation through the integration of machine learning. Recognizing the time-consuming nature of traditional simulation methods and the potential for machine learning to revolutionize architectural exploration, we set out on a journey to explore the future possibilities. Our work serves as an initial step in this direction, providing insights and promising results.

Our machine learning-based performance prediction model, with a particular focus on the Random Forest algorithm, has demonstrated that it is able to accurately estimate "instructions per cycle" (IPC). The alignment between predicted and actual IPC values, coupled with a low Root Mean Square Error (RMSE), validates the potential of our approach. Through feature importance analysis, we have revealed that cache sizes play a pivotal role in influencing application performance.

While we are encouraged by the promising results and insights gained in this preliminary exploration, we acknowledge that our work represents only the initial steps in the integration of machine learning and computer architecture simulation. As we conclude this paper, we emphasize its provisional nature and its place in the ongoing pursuit of more efficient and insightful architectural exploration.


\bibliographystyle{ACM-Reference-Format}
\bibliography{main}

\end{document}